\documentclass[aps,pre,amsmath,amssymb,groupedaddress]{revtex4}
\usepackage{graphicx}
\usepackage{color}

\newcommand{\re}[1]{(\ref{#1})}
\newcommand{\ci}[1]{\cite{#1}}
\newcommand{\be}{\begin{equation}}
\newcommand{\ee}{\end{equation}}
\newcommand{\ba}{\begin{array}{l}}
\newcommand{\ea}{\end{array}}
\newcommand{\lab}[1]{\label{#1}}
\renewcommand{\baselinestretch}{1.2}
\begin{document}
%\large

\title{Time-dependent quantum graph}
\author{D.U. Matrasulov$^{1,3}$\thanks{E-mail: d.matrasulov@polito.uz}, J.R. Yusupov$^1$, K.K. Sabirov$^1$, Z.A. Sobirov$^{1,2}$}
\affiliation{$^1$Turin Polytechnic University in Tashkent, 17.
Niyazov Str., 100095, Tashkent, Uzbekistan\\
$^2$Tashkent Financial Institute, 60A, Amir Temur Str., 100000,
Tashkent, Uzbekistan\\ $^3$Institute for Applied Physics, National University of Uzbekistan, 100174,  Tashkent, Uzbekistan}

\begin{abstract}
In this paper we study quantum star graphs with time-dependent bond lengths.
Quantum dynamics is treated by solving Schrodinger equation with time-dependent boundary conditions given on graphs.
Time-dependence of the average kinetic energy is analyzed. Space-time evolution of the Gaussian wave packet is treated for harmonically breathing star graph.

\end{abstract}

\maketitle

Quantum particle dynamics in nanoscale networks and discrete structures is of fundamental and practical importance.
Usually such systems are modeled by so-called quantum graphs, systems attracting much attention in physics \cite{Uzy1}-\cite{Uzy3} and
mathematics \cite{Exner1}-\cite{Exner3} during past two decades.

In physics quantum graphs were introduced as a 'toy' model for
studies of quantum chaos by Kottos and Smilansky \cite{Uzy1}.
However, the idea for studying of a system confined to a graph
dates back to Pauling \cite{Pauling} who suggested to use such systems for
modeling free electron motion in organic molecules. During last two decades quantum graphs found
numerous applications in modeling different discrete structures and networks in nanoscale and mesoscopic physics(e.g., see reviews
\cite{Uzy1}-\cite{Uzy3} and references therein).

Mathematical properties of the Schrodinger operators on graphs \cite{Exner1, Exner2, Exner3} inverse problems for quantum graphs \cite{Kurasov, Cheon}
were also subject for extensive research recently. Also, an experimental realization of quantum graphs was discussed earlier in the Ref. \cite{Hul}.
Despite the certain progress made in the study of quantum graphs some of important aspects are still remaining as
less- or not explored. Especially, this concerns the problems of
driven graphs, i.e. graphs perturbed by time-dependent external
forces. An important example of such a driving force is  that
caused by driven (moving) boundaries. Treatment of such system can
be reduced to the Schr\"odinger equation with time-dependent
boundary conditions. Earlier, the problem of time-dependent
boundary conditions has attracted much attention in the context of
quantum Fermi acceleration \cite{jose}-\cite{seb90}, though
different aspects of the problem was treated by many authors
\cite{mun81}-\cite{glas08}.
Detailed study of the problem can be found in series of papers by Makowski and co-authors \cite{mak91}-\cite{mak923}.
It was pointed out in the above Refs. that the problem of 1D box with the moving wall can be mapped onto that of time-dependent harmonic oscillator confined inside the static box \cite{mak91}.

In this paper we treat similar problem for quantum star graph, i.e. we study the problem of quantum graphs with time-dependent bonds.
In particular, we consider harmonically breathing quantum star graphs, the cases of contracting and expanding graphs. The latter can be solved exactly analytically. Motivation for the study of time-dependent graphs comes from such practically important problems as quantum Fermi acceleration in nanoscale network structures, tunable particle transport in quantum wire networks, molecular wires, different lattices and discrete structures. In particular, sites, vertices, nodes of such discrete structures can fluctuate that makes them time-dependent.  We will study time-dependence of the average kinetic energy and wave packet dynamics in harmonically breathing graphs.

Graphs are the systems consisting of  bonds which are connected at the  vertices. The bonds are connected according to a rule which is called topology of a graph. Topology of a graph is given by in terms of so-called adjacency matrix \cite{Uzy1,Uzy2}:
\begin{equation}
C_{ij}=C_{ji}=\left\{\begin{array}{c}1\qquad $if$\; i\; and\;
j\;
$are connected$\qquad\qquad\quad\\0\qquad $ otherwise$\end{array}\right.i,j=1,2,...,V.\nonumber
\end{equation}

Quantum dynamics of a particle in a graph is described in terms of
one-dimensional Schr\"odinger equation \ci{Uzy1,Uzy2} (in the units $\hbar =2m=1$):
\begin{align}
-i\frac{d^2\Psi_b(x)}{dx^2}=k^2\Psi_b(x),\quad b=(i,j),
\label{eq1}
\end{align}
where $b$denotes a bond connecting $i$th and $j$the vertices, and for each bond $b$, the component
$\Psi_b$ of the total wavefunction $\Psi_b$ is a solution of the
eq.\re{eq1}.

The wavefunction, $\Psi_b$, satisfies boundary conditions  at the vertices, which
ensure continuity and current conservation \ci{Uzy1}. General scheme for finding of eigenfunctions and eigenvalues
for such boundary conditions can be found in the Ref.\ci{Uzy1}. Different types of boundary conditions for the Schrodinger equation on graphs are discussed in the Refs.\cite{Exner1, Exner2, Exner3}. In the following we restrict our consideration by most simplest graph, so-called star graph.
The star graph consist of three or more bonds connected at the single vertex which is called branching point (see Fig.1).
Other ones are called edge vertices. The eigenvalue problem for a star graph with $N$ bonds is given by the following
Schr\"odinger equation:
$$
-\frac{d^2}{dx^2}\phi_j(y)=k^2\phi_j(y),\ \ 0\leq y \leq l_j, \ j=1,...,N.
$$
Here we consider the following  boundary conditions \cite{Keating}:
\begin{equation}\label{BC}
\left\{
\begin{array}{l}
  \phi_1|_{y=0}=\phi_2|_{y=0}=...=\phi_N|_{y=0}, \\
  \phi_1|_{y=l_1}=\phi_2|_{y=l_2}=...=\phi_N|_{y=l_N}=0, \\
  \sum\limits_{j=1}^N{\frac{\partial}{\partial
y}\phi_j|_{y=0}}=0. \\
\end{array}
\right.
\end{equation}

The eigenvalues can be found by solving the following equation \cite{Keating}
$$
\sum\limits^N_{j=1}{ctan(k_n l_j)}=0
$$
where corresponding eigenfunctions are given as \cite{Keating}

$$
\phi_j^{(n)}=\frac{B_n}{sin(k_n l_j)}
sin(k_n (l_j-y))
$$

$$
B_n=\sqrt\frac{2}{\sum_j{\frac{l_j+sin(2 k_n l_j)}{sin ^2 (k_n l_j)}}}.
$$

Time-dependent graph implies that lengths of the bonds of a graph are time-varying, i.e., when $L_j$ is a function of time.
In this case particle dynamics in graph is described by the following time-dependent Schr\"odinger equation:
\begin{align}
i\frac{\partial}{\partial
t}\Psi_j(x,t)=-\frac{\partial^2}{\partial x^2}\Psi_j(x,t), \ \
0<x<L_j(t), \ j=1,...,N, \label{Schr1}
\end{align}
with $N$ being the number of bonds.

In the following we will consider the boundary conditions given by
$$
\left\{
\begin{array}{l}
  \Psi_1|_{x=0}=\Psi_2|_{x=0}=...=\Psi_N|_{x=0}, \\
  \Psi_1|_{x=L_1(t)}=\Psi_2|_{x=L_2(t)}=...=\Psi_N|_{x=L_N(t)}=0, \\
  \sum\limits_{j=1}^N{\frac{\partial}{\partial
x}\Psi_j|_{x=+0}}=0. \\
\end{array}
\right.
$$
These boundary conditions imply that only edge vertices of the graph are moving while center (branching point) is fixed.
Furthermore, we  assume that  $L_j(t)$ is given as $L_j(t)=l_jL(t)$, where $L(t)$ is a continuous function and  $l_j$ are the positive constants.
Then using the coordinate transformation
$$
y=\frac{x}{L(t)},
$$
Eq. \re{Schr1} can be rewritten as
\begin{align}
i\frac{\partial}{\partial
t}\Psi_j(y,t)=-\frac{1}{L^2}\frac{\partial^2}{\partial
y^2}\Psi_j(y,t) + i\frac{\dot{L}}{L} y \frac{\partial}{\partial
y}\Psi_j(y,t), \ \ 0<y<l_j, \ j=1,...,N. \label{Schr2}
\end{align}
It is clear that the Schr\"odinger  operator in the right hand
side of Eq. \re{Schr2} is not Hermitian due to the presence of
second term. Therefore using the transformation
$$
\Psi_j(y,t)=\frac{1}{\sqrt{L}}e^{i\frac{L
\dot{L}}{4}y^2}\varphi_j(y,t),
$$
we can make it Hermitian as
\begin{align}
i\frac{\partial}{\partial
t}\varphi_j(y,t)=-\frac{1}{L^2}\frac{\partial^2}{\partial
y^2}\varphi_j(y,t) + \frac{L\ddot{L}}{4} y^2 \varphi_j(y,t), \ \
0<y<l_j, \ j=1,...,N. \label{Schr3}
\end{align}
We note that  the above transformations of the wave function remain the boundary conditions unchanged.

Time and coordinate variables in Eq. (\ref{Schr3}) can be separated only  in case when $L(t)$ obeys the equation
\begin{align}
\frac{L^3\ddot{L}}{4}=-C^2=const,
\label{length}
\end{align}
In this case using the substitution $\varphi_j(y,t)=\phi_j(y)\cdot exp\left(-i k^2 \int\limits_0^t\frac{ds}{L^2(s)}\right)$, we get
\begin{align}\label{eq for phi(y)}
\frac{d^2\phi_j}{dy^2}+(\lambda -C^2y^2)\phi_j=0, \ \ y\in
(0,l_j).
\end{align}

For  $C\not =0$ from Eq. (\ref{length}) we have
\be
L(t)=\sqrt{\alpha t^2+\beta t+\gamma},\ \  C^2=
\frac{1}{16}(\beta^2-4\alpha\gamma),
\ee
and
\be
L(t)=\sqrt{\beta t+\gamma},\ \  C^2=
\frac{1}{16}\beta^2.
\lab{equa1}
\ee

In both cases  exact solutions of Eq.\re{Schr3} can be obtained  in terms confluent hypergeometric functions.
In particular, for the case when time-dependence of $L(t)$ is given by Eq.\re{equa1} fundamental solutions of Eq.\re{Schr3} can be written as
\begin{align}
\phi_{j,1}=y\exp\left(\frac{C}{2}y\right)M\left(\frac{3}{4}-\frac{k}{4C},
\frac{3}{2}, -C  y^2\right),\nonumber
\end{align}
and
\begin{align}
\phi_{j,1}=\exp\left(\frac{C}{2}y\right)M\left(\frac{1}{4}-\frac{k}{4C},
\frac{1}{2}, -C y^2\right).\nonumber
\end{align}

Therefore the general solution of Eq.\re{Schr3} is given as
\begin{align}\label{gs phi}
\phi_j(y)=A_j\phi_{j,1}+B_j\phi_{j,2}.
\end{align}

From the boundary conditions given by Eq.(\ref{BC}) we have
\begin{align}
B_j=A, \ \ A_j=A\cdot\alpha_j(k),\ \ j=1,2,3,...,N,\nonumber
\end{align}
where $B$ is an arbitrary constant and
$$
\alpha_j(k)=-\frac{M\left(\frac{1}{4}-\frac{k}{4C},
\frac{1}{2}, -Cl_j^2\right)} {l_j
M\left(\frac{3}{4}-\frac{k}{4C}, \frac{3}{2},
-Cl_j^2\right)},\ \ j=1,2,...,N.
$$

Taking to account the  relations
$\left.\frac{d\phi_{j,1}(y)}{dy}\right|_{y=0}=1$,
$\left.\frac{d\phi_{j,2}(y)}{dy}\right|_{y=0}=\frac{C}{2},$
from Eq. (\ref{BC}) we obtain the following spectral equation for finding the eigenvalues, $k_n$ of Eq.\re{Schr3}:
\begin{align}\label{eigenvalue-eq(C not 0)}
\sum_{j=1}^N\frac{1}{l_j}\frac{M\left(\frac{1}{4}-\frac{k}{4C},
\frac{1}{2}, -Cl_j^2\right)}
{M\left(\frac{3}{4}-\frac{k}{4C}, \frac{3}{2},
-Cl_j^2\right)}=\frac{CN}{2}.
\end{align}

\begin{figure}[t]
\begin{center}
(a) \hskip 4cm (b)\\
\includegraphics[width=4cm, height=6cm, angle=-90]{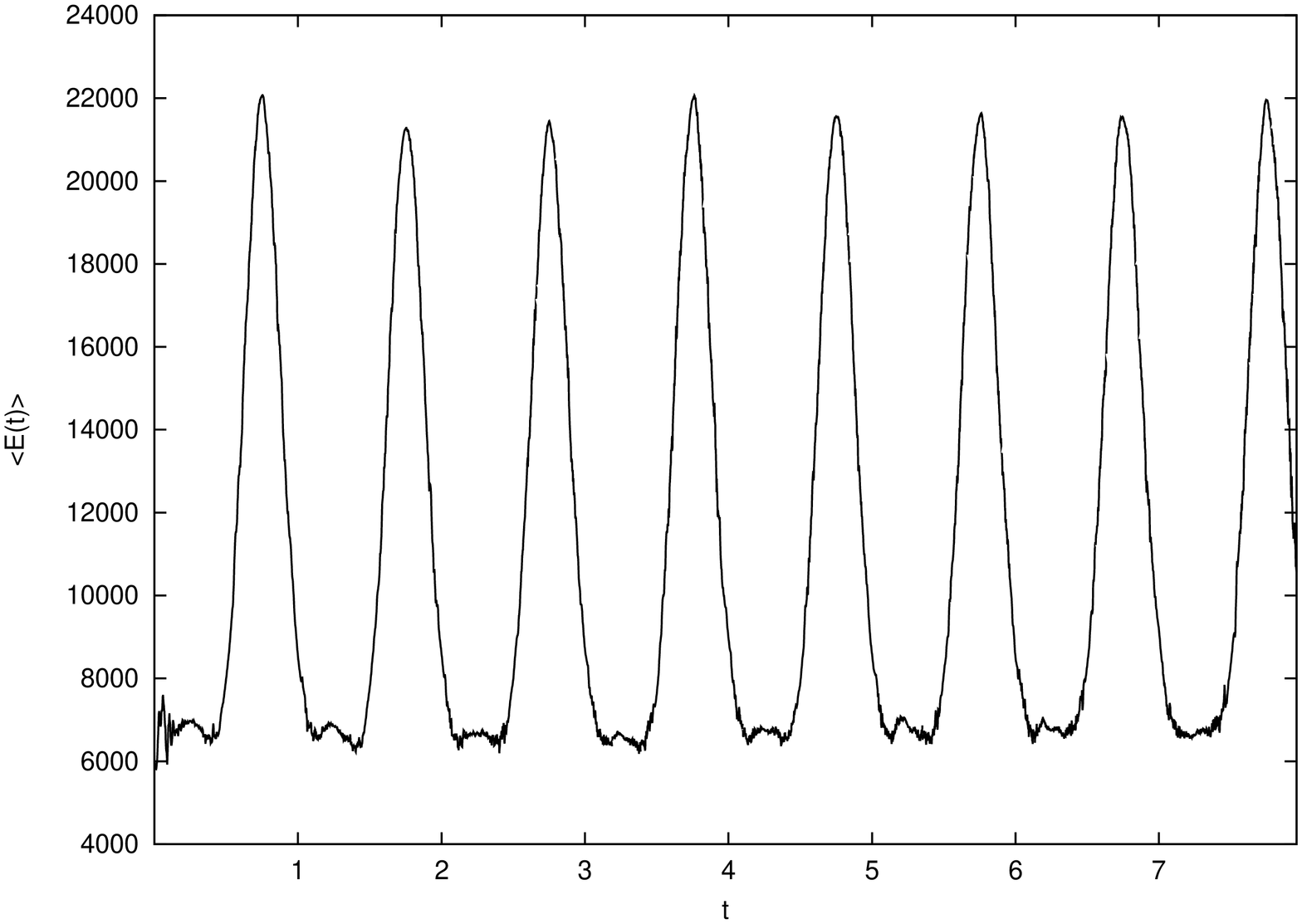}
\includegraphics[width=4cm, height=6cm, angle=-90]{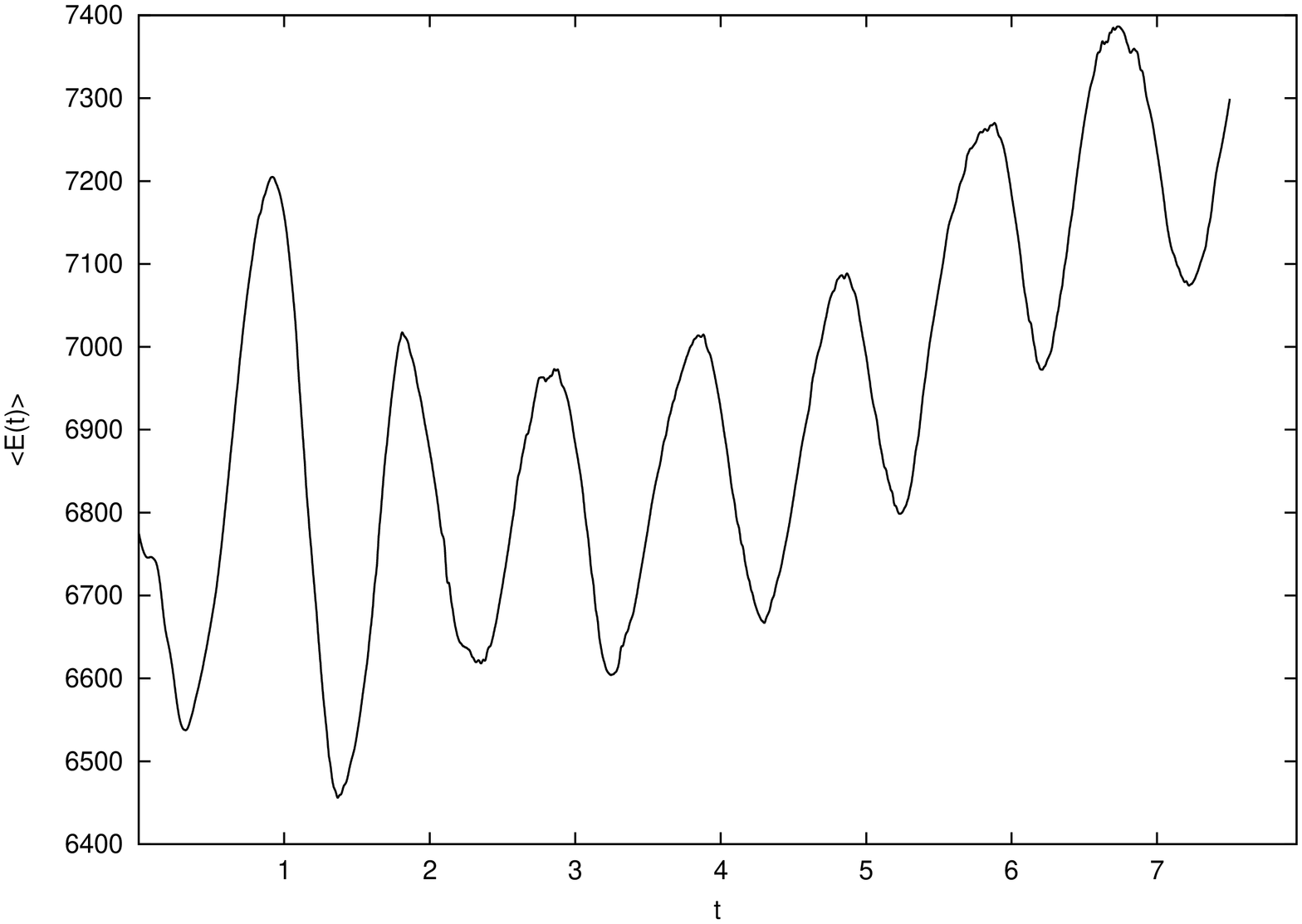}
\end{center}
\begin{center}
(c)\\
\includegraphics[width=4cm, height=6cm, angle=-90]{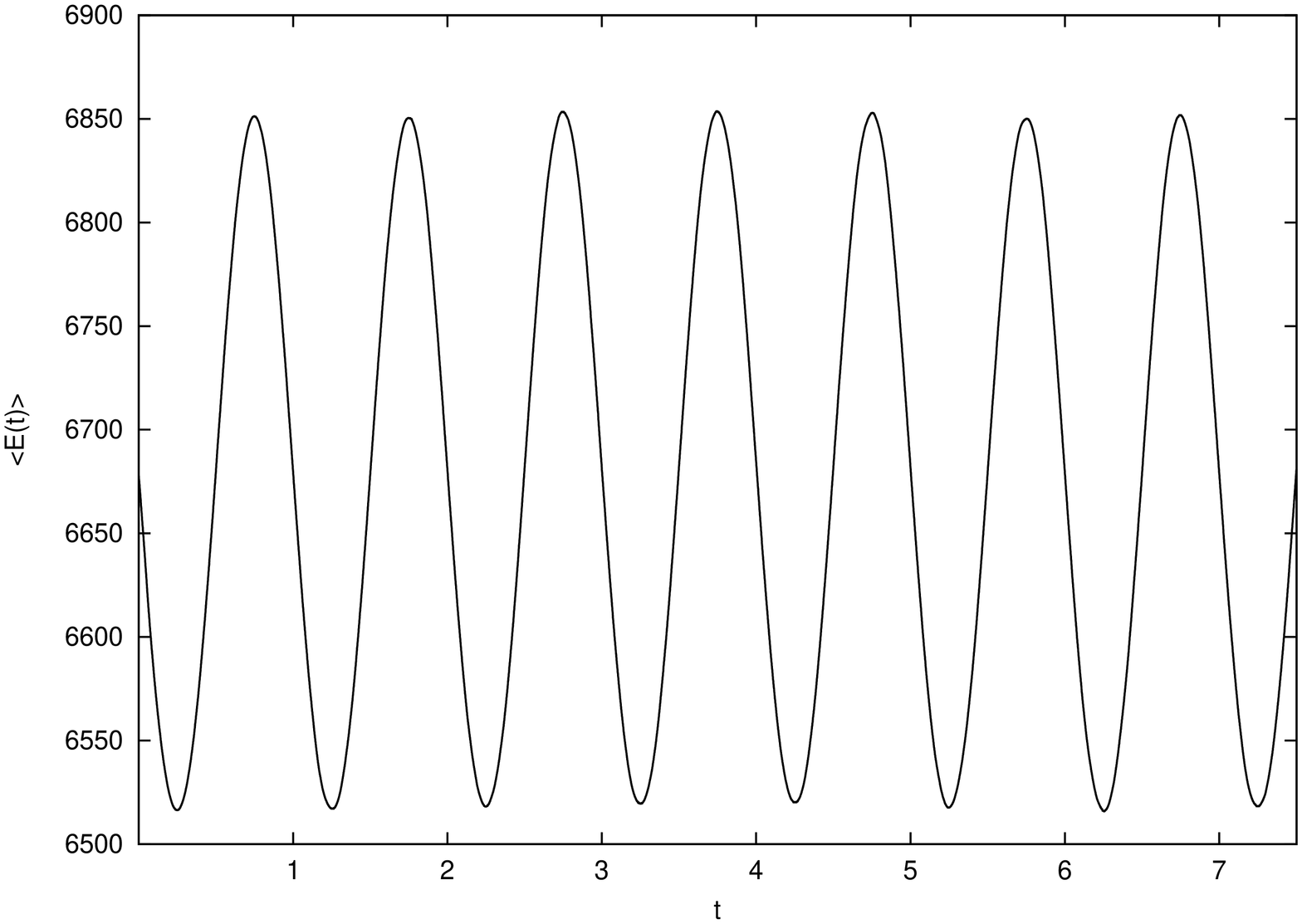}
\end{center}
\caption{
Time-dependence of the average kinetic energy for harmonically oscillating primary star graph.}
\label{fig:1}
\end{figure}

Thus the eigenfunctions of Eq.\re{Schr3} can be written as
\begin{align}
\phi_{j}(y,k_n)=A\left[\alpha_j(k_n)\phi_{j,1}(y)+\phi_{j,2}(y)\right],
\ \ j=1,2,...,N.
\end{align}

Furthermore, we provide the solution of Eq.\re{Schr3} for simplest case $L(t)=at+b$, which correspond to $C=0$ in Eq.\re{length}. In this case of the eigenvalues of Eq.\re{Schr3} can be written as
time-dependence of the wall is given as
\begin{align}\label{sol (C=0)}
\phi_j(y,k_n)=\frac{A}{sin(\sqrt{k_n} l_j)}
sin(\sqrt{k_n} (l_j-y)) , \ \ j=1,2,...,N,
\end{align}
where $k_n$ is the $n$th positive root of the equation
\begin{align}\label{eigenvalue-eq(C = 0)}
\sum_{j=1}^N ctan(l_j^2\sqrt{k})=0.
\end{align}
and  $L(t)>0$, $B$ is the normalization constant.

Now let us consider harmonically breathing graph, i.e. the case when time-dependence of $L(t)$  is given as
$$L(t)=b +a\cos\omega t$$
with $\omega =\frac{2\pi}{T}$ being oscillation frequency and T is the oscillation period. It is clear that in this case time and coordinate variables in Eq. \re{Schr3} cannot be separated.
Expanding $\varphi (y,t)$ in Eq. \re{Schr3} in terms of static graphs wave functions as
\be
\varphi_j(y,t)=\sum\limits_n{C_n(t) \phi_j^{(n)}(y)},
\lab{expand}
\ee
and inserting this expansion into Eq.\re{Schr3}we have
$$
\dot{C}_n(t)=\sum\limits_m{M_{mn}C_m(t)}
$$
where
$$
M_{mn}=-i\frac{k^2_m}{L^2(t)}-i\frac{L\ddot{L}}{4}K_{mn}
$$

The quantity we are interested to compute is the average kinetic energy which is defined as
\begin{align}\label{Def Energy}
E(t)=\langle\psi|H|\psi\rangle=\sum_{j=1}^N\int\limits_0^{L_j(t)}\left|\frac{\partial
\psi_j(x,t)}{\partial x}\right|^2dx.
\end{align}

In Fig. 1 time dependence of the average kinetic energy of the harmonically breathing star graph
is presented for different values of the breathing frequency and amplitude. As it can be seen from these plots, $<E(t)>$ is almost periodic for
$\omega =0.5$ and $a=1$, while for $\omega =10$ and $a=1$ such a periodicity is completely broken and energy grows in time.
For $\omega =10$ and $a=20$ the behavior of  $<E(t)>$ demonstrates "quasiperiodic behavior".  Appearing of periodic behavior in  $<E(t)>$ can be explained by synchronization of the motion of particle with the frequency. The lack of such synchronization causes breaking of the periodicity of the average energy in time.

Furthermore, we consider wave packet evolution in harmonically breathing star graph by taking the wave function at $t = 0$ (for the first bond) as the following Gaussian wave packet:
\be
\Psi_1(x,0) = \Phi(x) = \frac{1}{\sqrt{2\pi}\sigma}e^{-\frac{(x-\mu)^2}{2\sigma}},
\lab{wp}
\ee
with $\sigma$ being the width of the packet.
For other bonds initial wave function is assumed to be zero, i.e. $\Psi_2(x,0) =\Psi_3(x,0) =0$. Then for the
initial values of the functions $\varphi^{(j)}(y,t)$ in Eq. \re{expand} we have
$$
\varphi^{(j)}(y,0) =L(0)e^{-i\frac{L(0)\dot L(0)}{4}y^2}\Phi(y).
$$
Correspondingly, the expansion coefficients at $t=0$ can be written as
$$
C_n(0) = \sum_j\int_0^{l_j} \varphi^{(j)}(y,0) \phi_n^{(j)}(y)^*.
$$

In calculation of the wave packet evolution we will choose initial condition as the wave packet being on the first bond only, while for other two bonds the wave function at $t = 0$
is taken as zero.
In Fig.2 the time evolution of the wave packet is plotted for harmonically breathing primary star graph whose bonds oscillate according to the law $L(t)=40 +a\cos\omega t$ .
The oscillation parameters (frequency and amplitude) are chosen as follows:
a) $\omega =10,$\; $a=20;$\;b) $\omega =10,$\; $a=1;$\; c) $\omega =0.5,$\; $a=1.$  Fig.2d presents wave packet evolution in static(time-independent) star graph.
At $t=0$ a Gaussian packet of the width $\sigma$  and velocity $v_0$ is assumed being in the first bond.
As it can be seen from these plots, for higher frequencies dispersion of the packet and its transition  to other bonds occur more faster compared to that for smaller smaller frequencies.
Again, an important role plays here possible synchronization between the bond edge  and wave packets motions. Existence or absence of ssuch synchronization defines how the collision of
the packet with the bond edges will occur and how extensively it gains or loses its energy. Therefore more detailed treatment of the wave packet dynamics in harmonically breathing
graphs should be based on the  analysis of the role of synchronization and its criterions.

In this paper we have treated  time-dependent quantum network by considering expanding and harmonically breathing quantum star graphs.
Edge boundaries are considered to be time-dependent, while branching point is assumed to be fixed(static).
Time-dependence of the average kinetic energy and space-time evolution of the Gaussian wave packet are studied by solving the Schrodinger equation with time-dependent boundary conditions.
It is shown that for certain frequencies  energy is a periodic function of time, while for others it gan be non-monotonically growing function of time. Such a feature can be caused
by possible  synchronization of of the particles motion and the motions of the moving edges of graph bonds. Similar feature can be seen also from the analysis of the wave packet evolution.
The above study can be useful for the treatment of particle transport in different discrete structures, such as molecular and quantum wire networks, networks of carbon nanotubes, crystal lattices,  and others nanoscale systems
that can be modeled by quantum graphs.

\begin{figure}[t]
\begin{center}
(a) \hskip 4cm (b)\\
\includegraphics[width=4cm, height=6cm]{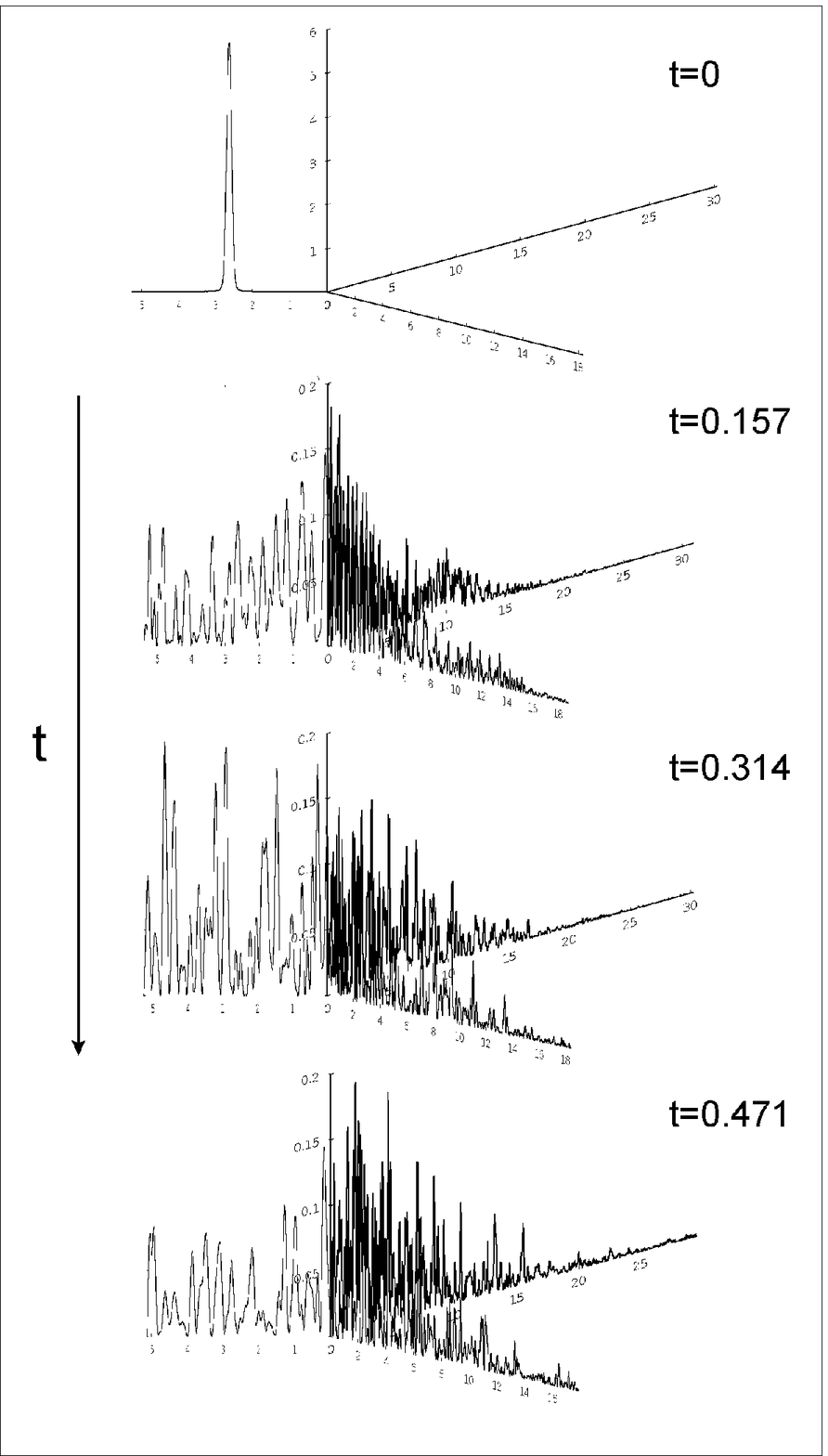} \hskip 1cm
\includegraphics[width=4cm, height=6cm]{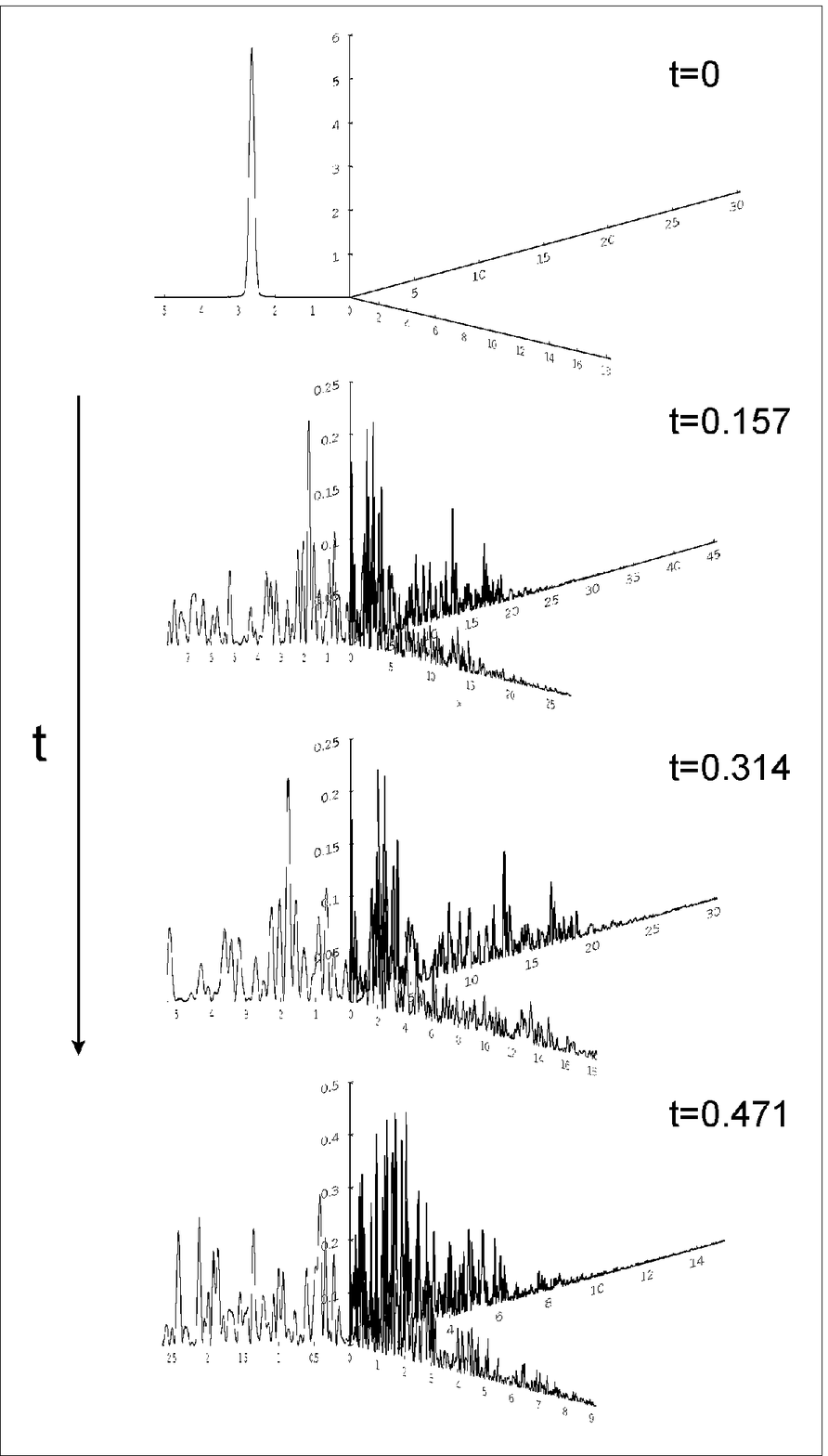}
\end{center}
\begin{center}
(c) \hskip 4cm (d)\\
\includegraphics[width=4cm, height=6cm]{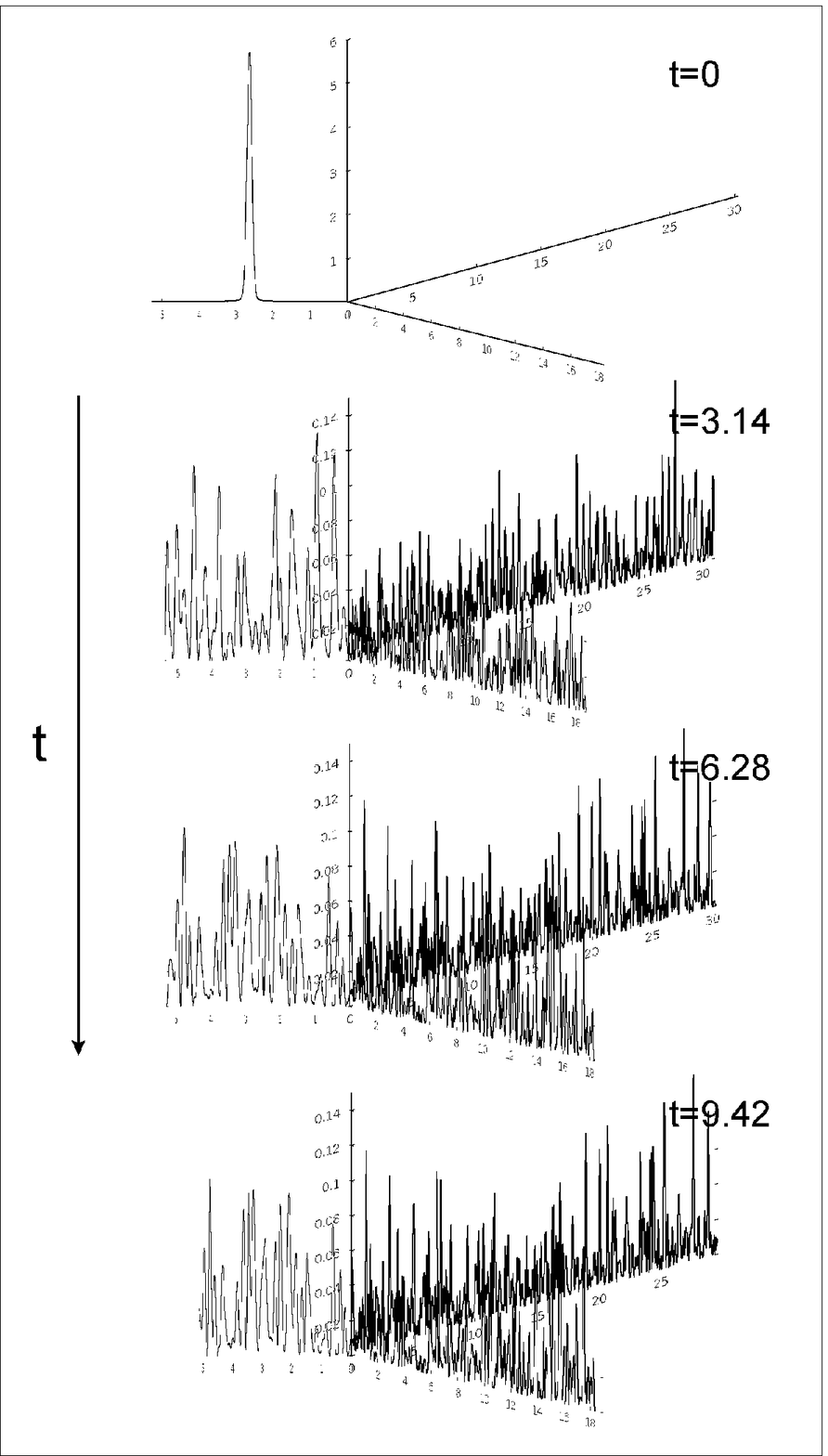} \hskip 1cm
\includegraphics[width=4cm, height=6cm]{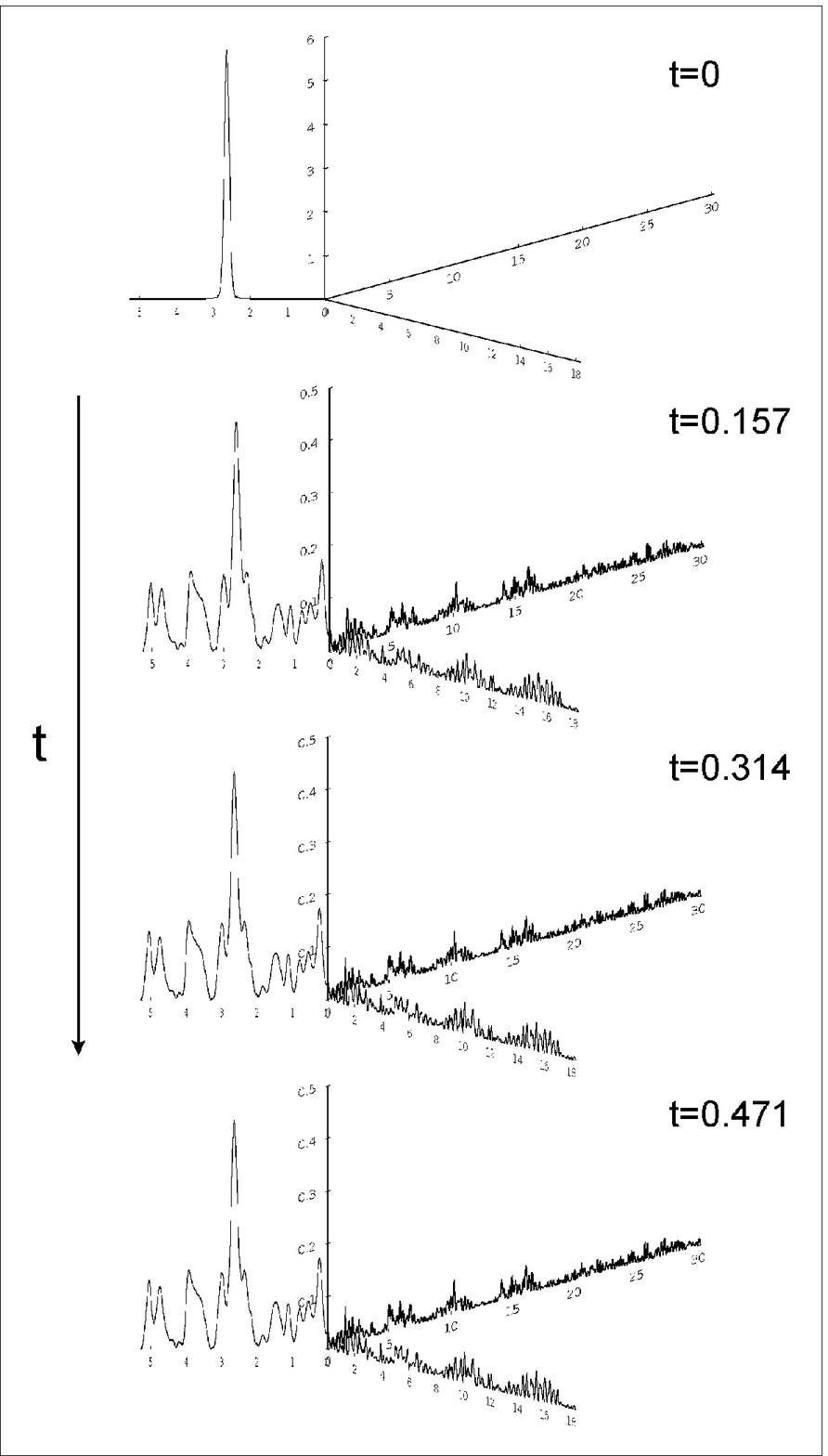}
\end{center}
\caption{ Time evolution of the Gaussian wave packet given by Eq. \re{wp} for the parameters:
a) $\omega =10,$\; $a=20;$\;b) $\omega =10,$\; $a=1;$\; c) $\omega =0.5,$\; $a=1;$\; d) Wave  packet evolution in static star graph.}
\label{fig:2}
\end{figure}


\begin{thebibliography}{99}
\bibitem{Uzy1} Tsampikos Kottos and Uzy Smilansky, Ann.Phys., {\bf 76} 274 (1999).
\bibitem{Uzy2} Sven Gnutzmann and Uzy Smilansky, Adv.Phys. {\bf 55} 527 (2006).
\bibitem{Uzy3} S. GnutzmannJ.P. Keating b, F. Piotet, Ann.Phys., {\bf 325} 2595 (2010).
\bibitem{Pauling} L. Pauling, J. Chem. Phys. {\bf 4} 673 (1936).
\bibitem{Exner1} P.Exner, P.Seba, P.Stovicek, J. Phys. A: Math. Gen. {\bf 21} 4009-4019 (1988).
\bibitem{Exner2} P.Exner, P.Seba,  Rep. Math. Phys., {\bf 28} 7 (1989).
\bibitem{Exner3} P.Exner, Ann. Inst. H. Poincare: Phys. Theor, {\bf 66} 359 (1997).
\bibitem{Kurasov} J. Boman, P. Kurasov, Adv. Appl. Math., {\bf 35}, 58 (2005)
\bibitem{Cheon}, T. Cheon, P. Exner, O. Turek, Ann.Phys., {\bf 325} 548 (2010).
\bibitem{Hul} Oleh Hul et al,  Phys. Rev. E {\bf 69}, 056205 (2004).
\bibitem{Keating} J.P.Keating, Contemp. Math.,   {\bf 415}, 191 (2006).
\bibitem{jose} J.V. Jose, R. Gordery Phys. Rev. Lett. {\bf 56}, 290 (1986).
\bibitem{karner} G. Karner, Lett. Math. Phys. A {\bf 17}, 329 (1989).
\bibitem{seb90} P. Seba. Phys. Rev. A {\bf 41}, 2306 (1990).

\bibitem{Doescher} S.W. Doescher and M.H. Rice Am. J. Phys. {\bf 37}, 1246 (1969).
\bibitem{mun81} A. Munier, J.R. Burgan, M. Feix and E. Fijalkow. J. Math. Phys. {\bf 22}, 1219 (1981).
\bibitem{pinder} D.N. Pinder, Am. J. Phys. {\bf 58}, 54 (1990).
\bibitem{razavy} M. Razavy. Phys. Rev. A {\bf 44}, 2384 (1991).
\bibitem{per} P. Pereshogin, P. Pronin, Phys. Lett. A {\bf 156}, 12 (1991).
\bibitem{sch91} C. Scheininger and M. Kleber. Physica D {\bf 50}, 391 (1991).
\bibitem{mak91} A.J. Makowski and S.T. Dembinski. Phys. Lett. A {\bf 154}, 217 (1991).
\bibitem{mak92} A.J. Makowski and P. Peplowski. Phys. Lett. A {\bf 163}, 142 (1992).
\bibitem{mak923} A.J. Makowski. J. Phys. A: Math. Gen. {\bf 25},  3419(1992).
\bibitem{will} J.E. Willemsen, Phys. Rev. E {\bf 50}, 3116 (1994).
\bibitem{mora} D.A. Moralez, Z. Parra, R. Almeida, Phys. Lett. A {\bf 185}, 273 (1994).
\bibitem{yuce} C. Yuce,  Phys. Lett. A {\bf 321}, 291 (2004).
\bibitem{glas08} M.L. Glasser, J. Mateo, J. Negro and L.M. Nieto,  Chaos, Solitons
and Fractals, {\bf 41}, 2067 (2009).

\end{thebibliography}
\end{document}